\documentclass[11pt]{wlscirep}
\usepackage{lineno}
\usepackage{amsmath}
\usepackage{amssymb}
\usepackage{nicefrac} 

\title{Inter-individual and inter-site neural code conversion without shared stimuli}
\author[1,2,3,*]{Haibao Wang}
\author[1]{Jun Kai Ho}
\author[1,2]{Fan L. Cheng}
\author[1,2]{Shuntaro C. Aoki}
\author[1]{Yusuke Muraki}
\author[1,2]{Misato Tanaka}
\author[1,2,*]{Yukiyasu Kamitani}
\affil[1]{Graduate School of Informatics, Kyoto University, Yoshida-honmachi, Sakyo-ku, Kyoto, 606-8501, Japan}
\affil[2]{Department of Neuroinformatics, ATR Computational Neuroscience Laboratories, Hikaridai, Seika, Soraku, Kyoto, 619-0288, Japan}
\affil[3]{Guardian Robot Project, RIKEN, Hikaridai, Seika, Soraku, Kyoto, 619-0288, Japan}

\affil[*]{Correspondence: haibaowa@gmail.com, kamitani@i.kyoto-u.ac.jp} 

\begin{abstract}
Inter-individual variability in fine-grained functional brain organization poses challenges for scalable data analysis and modeling. Functional alignment techniques can help mitigate these individual differences but typically require paired brain data with the same stimuli between individuals, which is often unavailable. We present a neural code conversion method that overcomes this constraint by optimizing conversion parameters based on the discrepancy between the stimulus contents represented by original and converted brain activity patterns. This approach, combined with hierarchical features of deep neural networks (DNNs) as latent content representations, achieves conversion accuracy comparable to methods using shared stimuli. The converted brain activity from a source subject can be accurately decoded using the target's pre-trained decoders, producing high-quality visual image reconstructions that rival within-individual decoding, even with data across different sites and limited training samples. Our approach offers a promising framework for scalable neural data analysis and modeling and a foundation for brain-to-brain communication.
\end{abstract}
\begin{document}

\flushbottom
\maketitle

\thispagestyle{empty}

\clearpage
\generateFig{figureSuggestions}{figure1.png}{0.85}
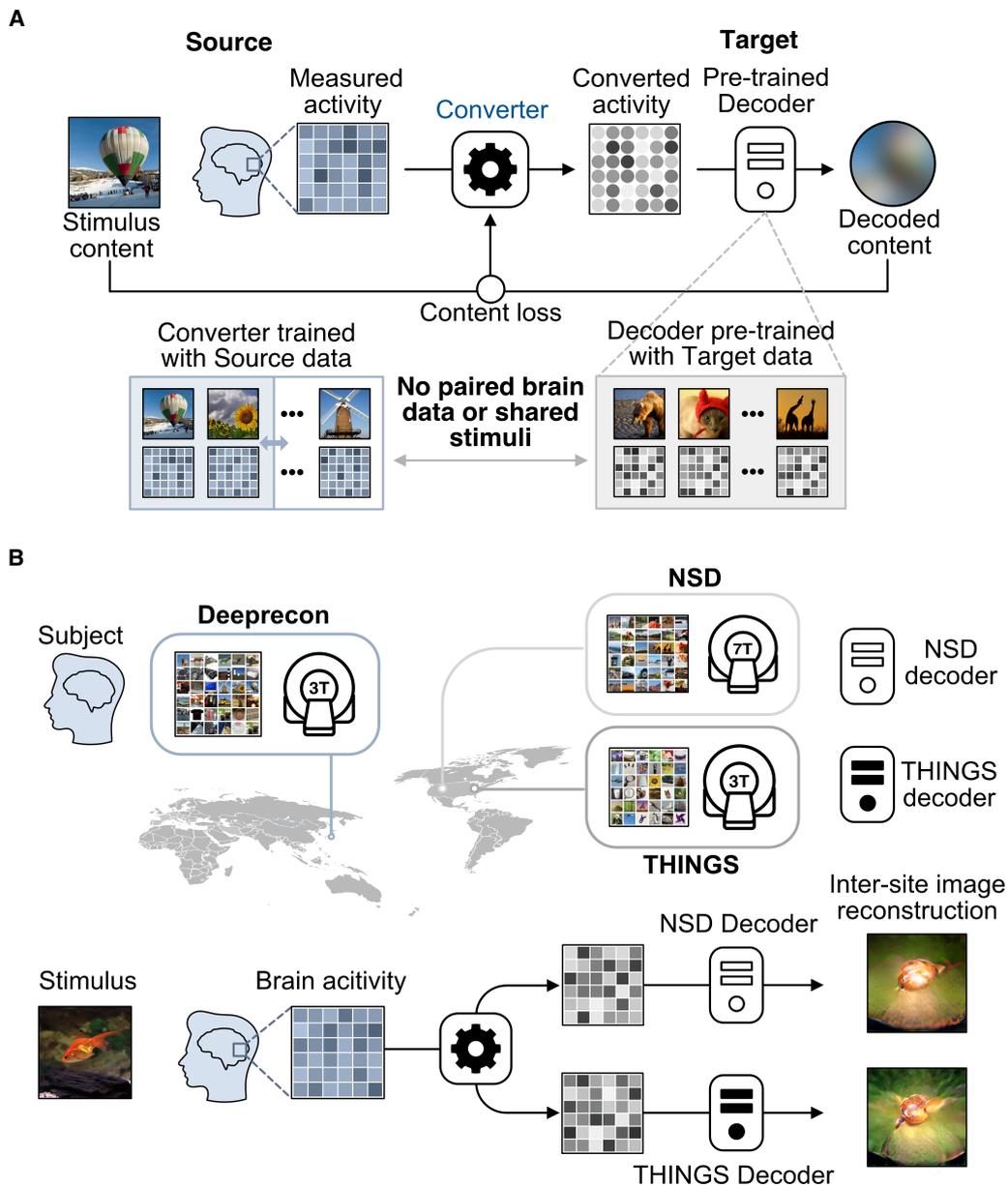
    {Content loss-based neural code conversion and inter-individual (-site) image reconstruction.} 
    {\textbf{(A)} Training the neural code converter with content loss-based optimization. The target subject's training data is used to pre-train the target decoder, whereas the source subject's training data is used for the converter training. The converter is optimized so that the converted brain activity is decoded into content representations that closely resemble that of the stimulus given to the source subject. During the training stage, there was no need for paired brain data or shared stimuli between source and target. 
        \textbf{(B)} The example of inter-individual (-site) visual image reconstruction. The trained converter functions across different datasets, which may have different stimuli and scanners with different resolutions. When presented with a novel stimulus, the subject’s brain activity pattern from one dataset is converted into the space of another dataset. The converted pattern is then decoded into image feature representations by a feature decoder within the target dataset, enabling inter-individual (-site) image reconstruction as perceived by the source subject. 
    }

\generateFig{figuretwo}{figure2.png}{1}
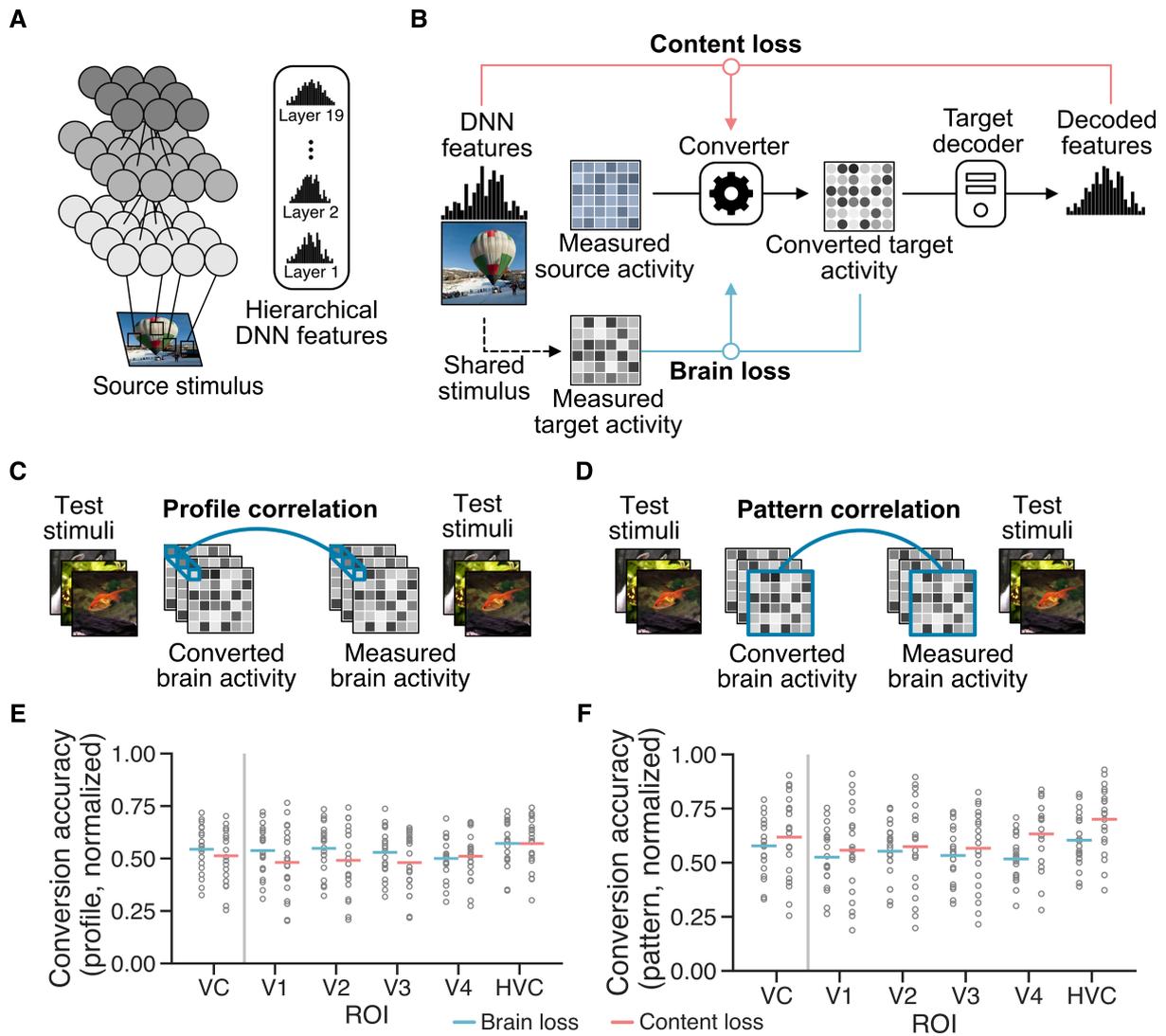
    {Neural code conversion using content loss and brain loss.} 
    {\textbf{(A)} Hierarchical DNN features as visual contents. \textbf{(B)} The overview of converter training using content loss-based optimization and brain loss-based optimization. The content loss-based optimization minimizes the loss between the DNN features decoded from the converted target activity and those extracted from the corresponding stimulus. The brain loss-based optimization minimizes the loss between the converted target activity and the measured target activity.
        \textbf{(C)} Evaluation of neural code conversion (profile). Pearson correlation coefficient between the sequences of converted and true voxel responses to the 50 test stimuli calculated for profile correlation.
        \textbf{(D)} Evaluation of neural code conversion (pattern). Pearson correlation coefficient between the converted and measured voxel patterns for a test stimulus was calculated for pattern correlation.
        \textbf{(E)} Conversion accuracies measured by profile correlation. Distributions of the profile correlation coefficients of 20 pairs are shown for the VC and visual subareas. The horizontal bars represent the mean accuracies across the 20 pairs, while each dot indicates an individual pair's mean correlation coefficient over stimuli.
        \textbf{(F)} Conversion accuracies measured by pattern correlation. The two converters are compared as in (E) but with pattern correlation coefficients.
    }

\generateFig{figurethree}{figure3.jpg}{0.82}
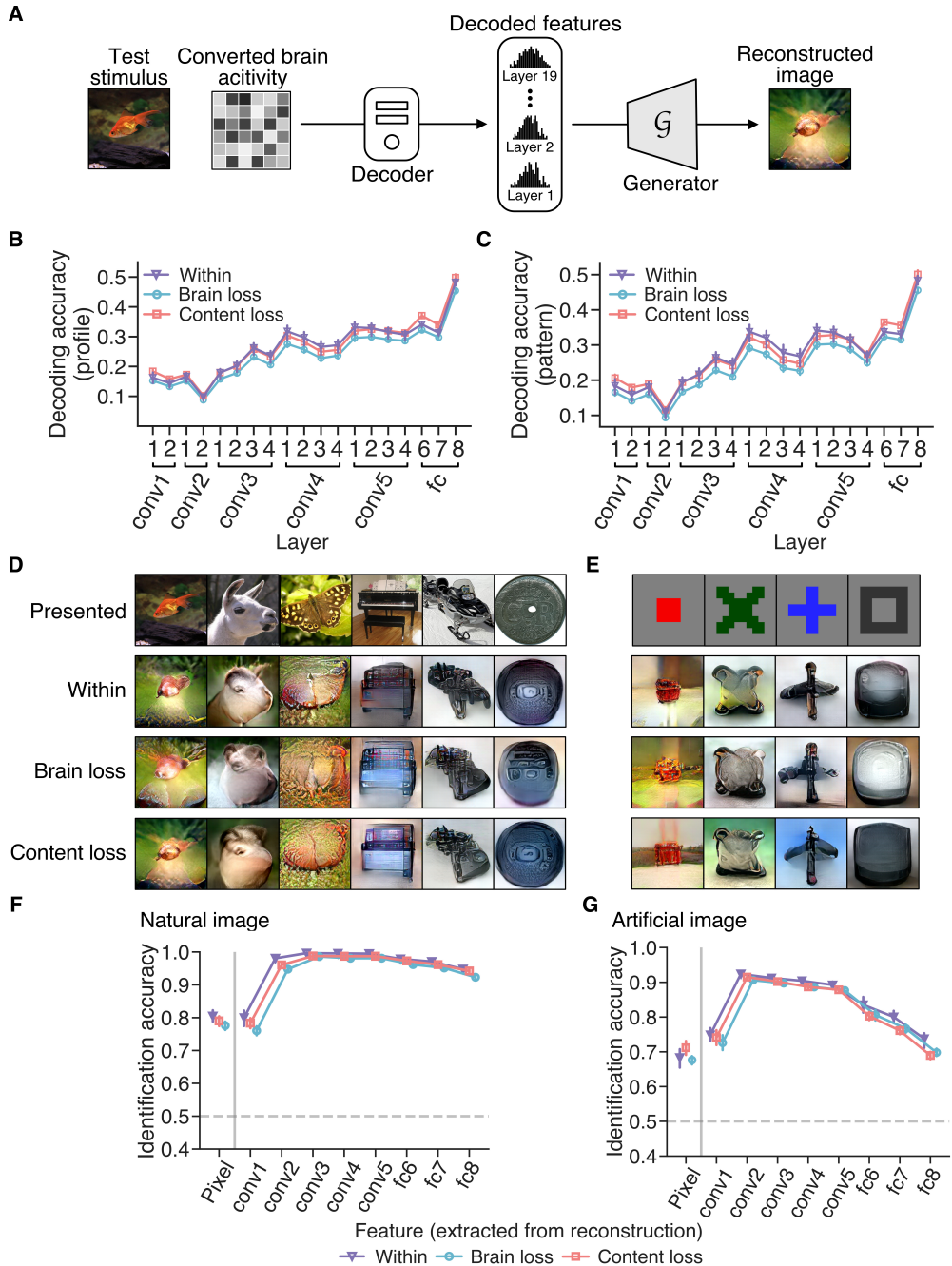
    {Inter-individual decoding and image reconstruction.} 
    {\textbf{(A)} DNN feature decoding and image reconstruction. The converted brain activity was decoded into DNN features using the decoder in the target space. The decoded features were fed into a reconstruction algorithm to reconstruct images.
        \textbf{(B)} Feature decoding accuracy measured by profile correlation. The mean profile correlation for each layer of the VGG19 model is shown for the Within, Brain loss, and Content loss conditions (VC; error bars, 95\% C.I. from five subjects for the Within condition, and from 20 individual pairs for the Brain loss and Content loss conditions).
        \textbf{(C)} Feature decoding accuracy measured by pattern correlation. The three conditions are compared as in (B) but with pattern correlation.
        \textbf{(D)} Image reconstructions of natural images. The reconstructions under the three analytical conditions for each stimulus image were all from the same source subject (VC; source: Subject 1, target: Subject 2; see Supplementary Fig.2 for the reconstructions across all individual pairs).
        \textbf{(E)} Image reconstructions of artificial images. The reconstructions under the three analytical methods were shown as in (D) but with artificial images (see Supplementary Fig.3 for the reconstructions across all individual pairs). 
        \textbf{(F)} Identification accuracy of natural images. The identification analysis was performed using the pixel values and the extracted DNN feature values of the reconstructions. The mean identification accuracy was calculated over all reconstructed images for each subject or individual pair. DNN features of images were extracted from the eight layers of the AlexNet model (error bars, 95\% C.I. from five subjects or 20 pairs; dashed lines, 50\% chance level).
         \textbf{(G)} Identification accuracy of artificial images.
    }

    \generateFig{figurefour}{figure4.png}{1}
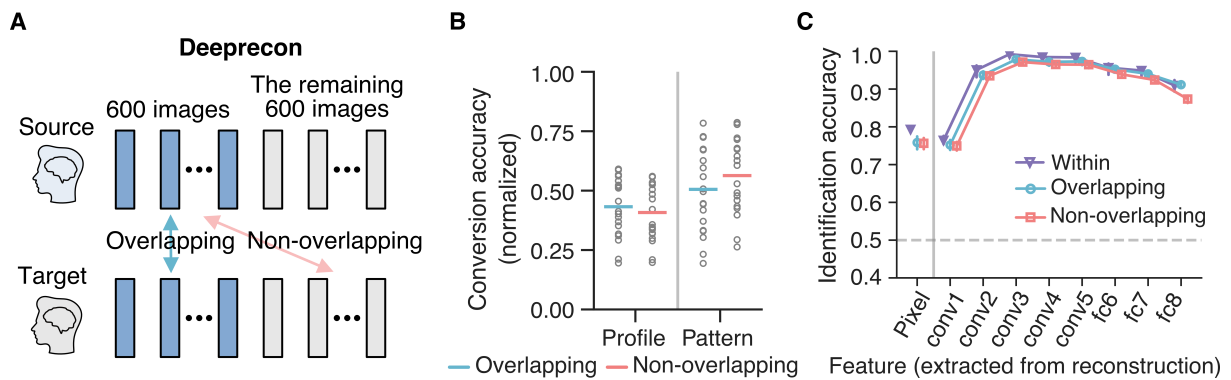
    {No need for overlapping (shared) stimuli between converter and decoder trainings.} 
    {\textbf{(A)} Non-overlapping stimuli for source and target subjects. The Deeprecon training samples were randomly split into two distinct halves. Source and target subjects were allocated different halves of training samples, ensuring no overlap in image stimuli between the source and target subjects.
    \textbf{(B)} Conversion accuracy. Distributions of the profile and pattern correlation coefficients of 20 individual pairs are shown for the VC (see Supplementary Fig.5A, B for conversion accuracies across different ROIs). The horizontal bars show the mean accuracies across 20 pairs, and each dot represents the mean correlation coefficient for an individual pair. The “Overlapping” indicates that stimuli are shared between converter and decoder trainings, while the “Non-overlapping” means no stimuli are shared. 
    \textbf{(C)} Identification accuracy. Pairwise identification was performed using the pixel values and the extracted DNN feature values from the reconstructions. The mean identification accuracy was calculated over all reconstructed images for each subject or individual pair (error bars, 95\% C.I. from 20 pairs; dashed lines, 50\% chance level; see Supplementary Fig.5E for the reconstructed images).
    }

    \generateFig{figurefive}{figure5.jpg}{1}
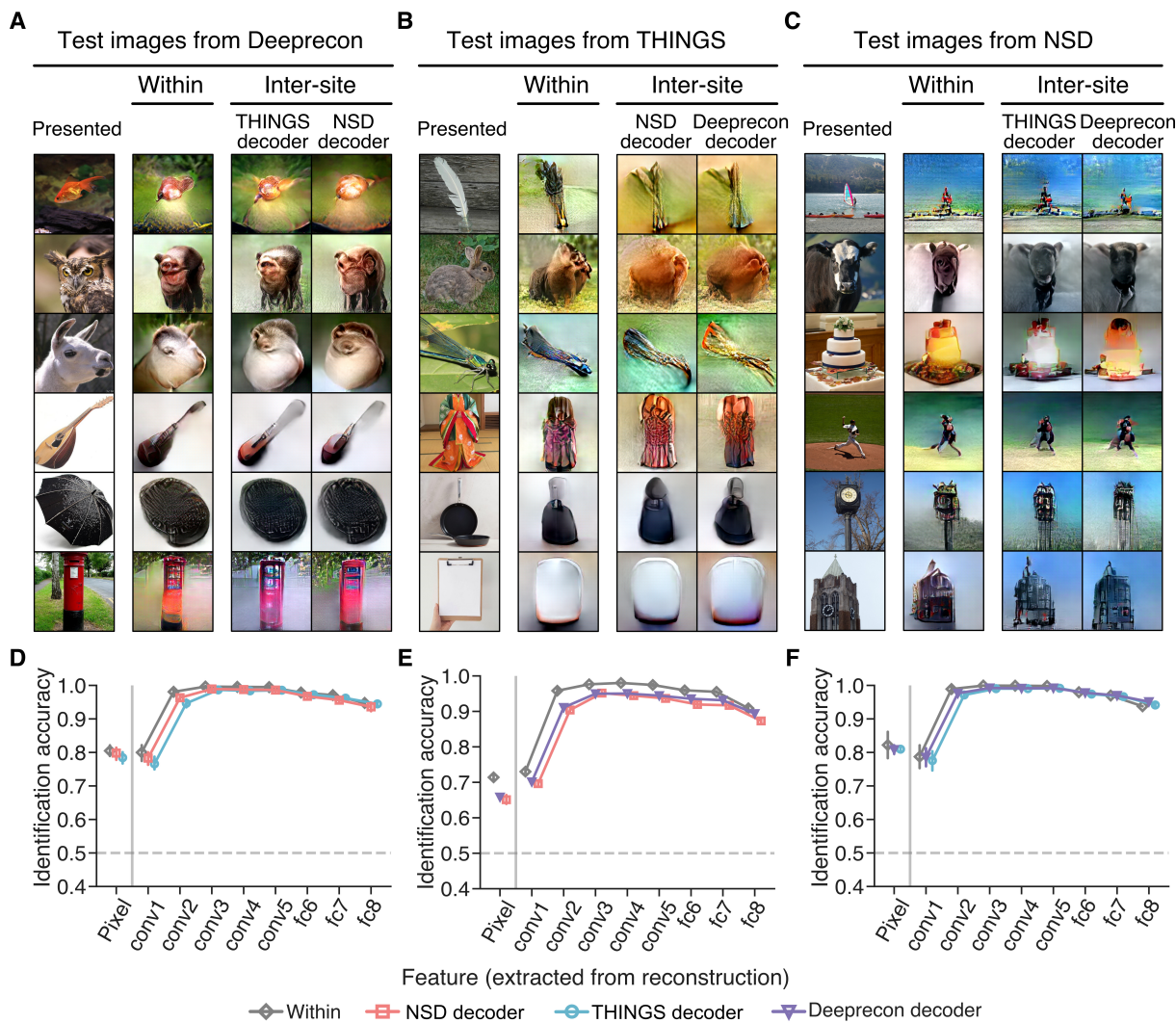
    {Inter-site image reconstruction.} 
    {\textbf{(A)} Reconstructions for Deeprecon test stimuli. Within reconstructions were obtained using decoders from the same subject, while inter-site reconstructions were obtained using the target’s decoders from other datasets (VC; source: Subject1, target: Subject 6, 8; see Supplementary Fig.7 for reconstructions of all pairs). 
    \textbf{(B)} Reconstructions for THINGS test stimuli. Examples for source Subject 6 and target Subjects 1 and 8 are shown (VC; see Supplementary Fig.8 for reconstructions of all pairs).
    \textbf{(C)} Reconstructions for NSD test stimuli. Examples for source Subject 8 and target Subjects 1 and 6 are shown (VC; see Supplementary Fig.9 for reconstructions of all pairs).
    \textbf{(D)} Identification accuracy for Deeprecon test stimuli. Pairwise identification was performed using the pixel values and the extracted DNN feature values (AlexNet) from the reconstructions. Mean identification accuracy was calculated over all reconstructed images for each subject or individual pair (error bars, 95\% C.I. from five subjects or 20 pairs; dashed lines, 50\% chance level).
    \textbf{(E)} Identification accuracy for THINGS test stimuli. The mean identification accuracy was calculated as in (D) but with test images from THINGS.
    \textbf{(F)} Identification accuracy for NSD test stimuli. The mean identification accuracy was calculated as in (D) but with test images from the NSD.}

    \generateFig{figuresix}{figure6.jpg}{0.92}
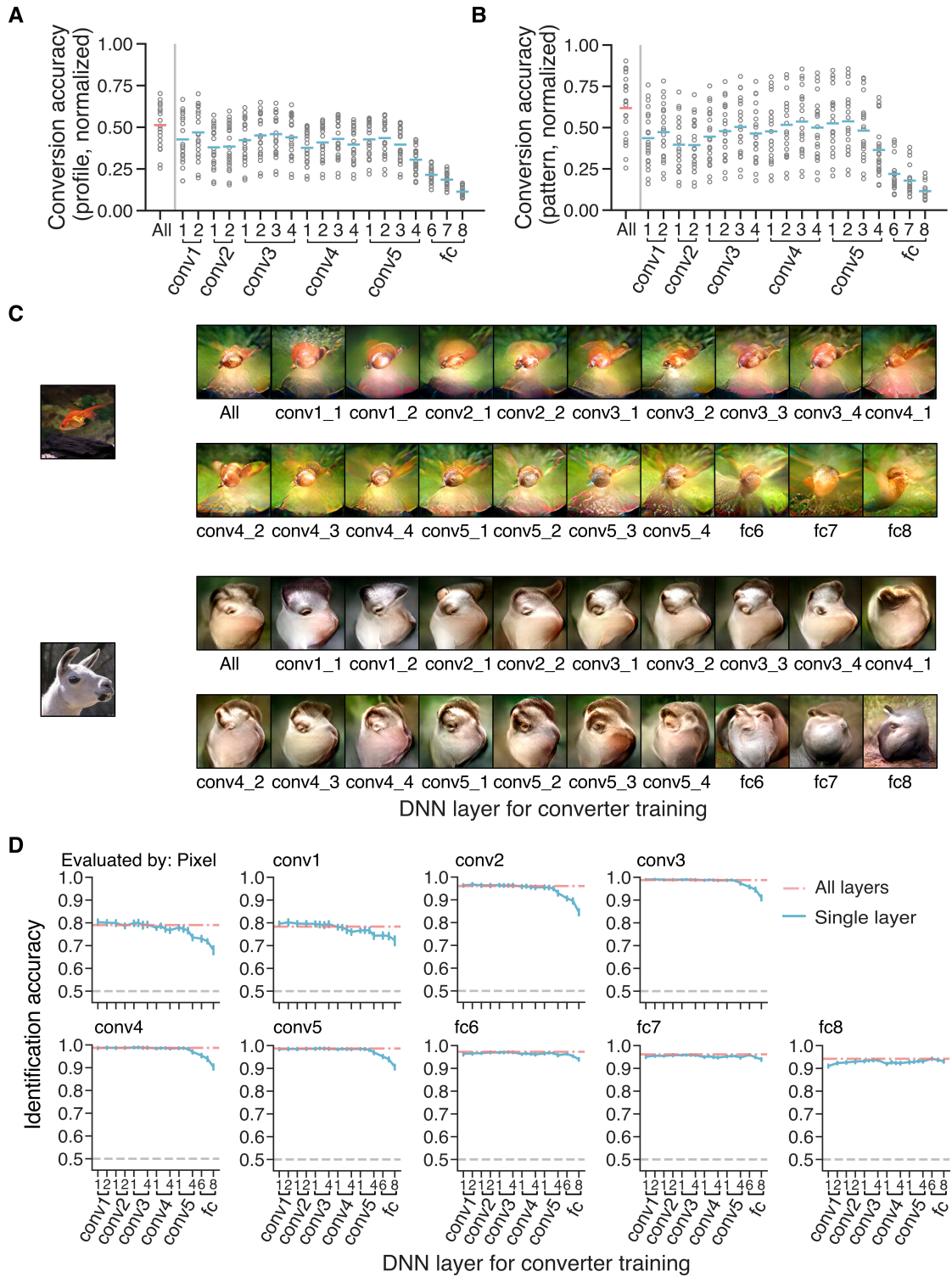
    {Converters trained using the loss from different DNN layers.} 
    {\textbf{(A)} Conversion accuracy measured by profile correlation. The profile correlation coefficients calculated for 20 individual pairs (VC) are used to compare converters trained using loss from different DNN layers. The horizontal bars represent the mean accuracies across the 20 pairs, while each dot indicates an individual pair's mean correlation coefficient over stimuli. The results are shown with the converter trained using loss from all layers (All).
    \textbf{(B)} Conversion accuracy measured by pattern correlation. The comparison is shown as in (A) but with pattern correlation coefficients.
    \textbf{(C)} Reconstructions. All reconstructed images are generated from the same pair (VC; source: Subject 1, target: Subject 2).
    \textbf{(D)} Pairwise identification was performed using the pixel values and the extracted DNN feature values (AlexNet) from the reconstructions. The “All layers” refers to converters trained using the loss from all DNN layers, while the “single layer” corresponds to converters trained using the loss from single DNN layers. The mean accuracy was calculated in each subject and condition (error bars, 95\% C.I. from 20 pairs; dashed lines, 50\% chance level).
    }

    \generateFig{figureseven}{figure7.png}{1}
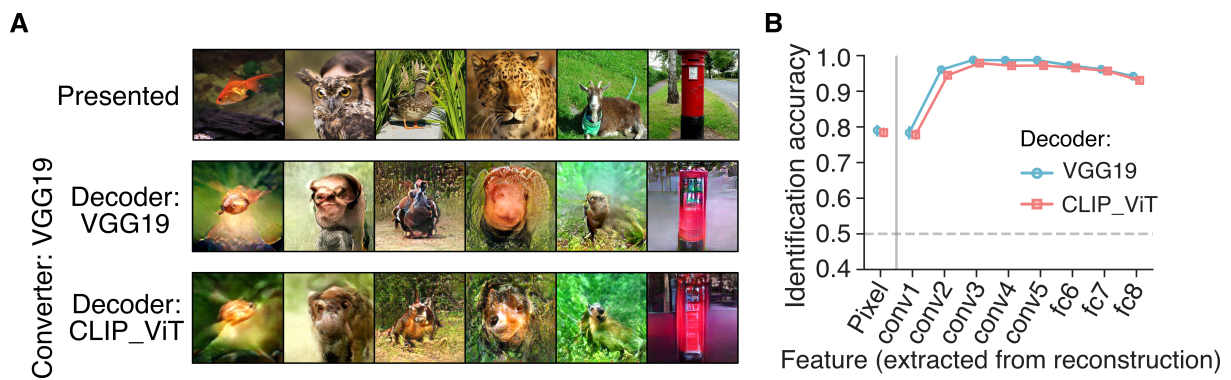
    {Reconstruction by converter and decoder trained with different DNNs.} 
    {\textbf{(A)} Reconstructed images using the VGG19 decoder and CLIP\_ViT decoder. The reconstructions were generated from the same pair (VC; source: Subject 1, target: Subject 2; see Supplementary Fig.13 for reconstructions of all pairs). \textbf{(B)} Identification accuracy. Pairwise identification was performed using the pixel values and the extracted DNN feature values (AlexNet) from the reconstructions obtained via the VGG19 decoder and CLIP\_ViT decoder (error bars, 95\% C.I. from 20 pairs; dashed lines, 50\% chance level). }

    \generateFig{figureeight}{figure8.jpg}{1}
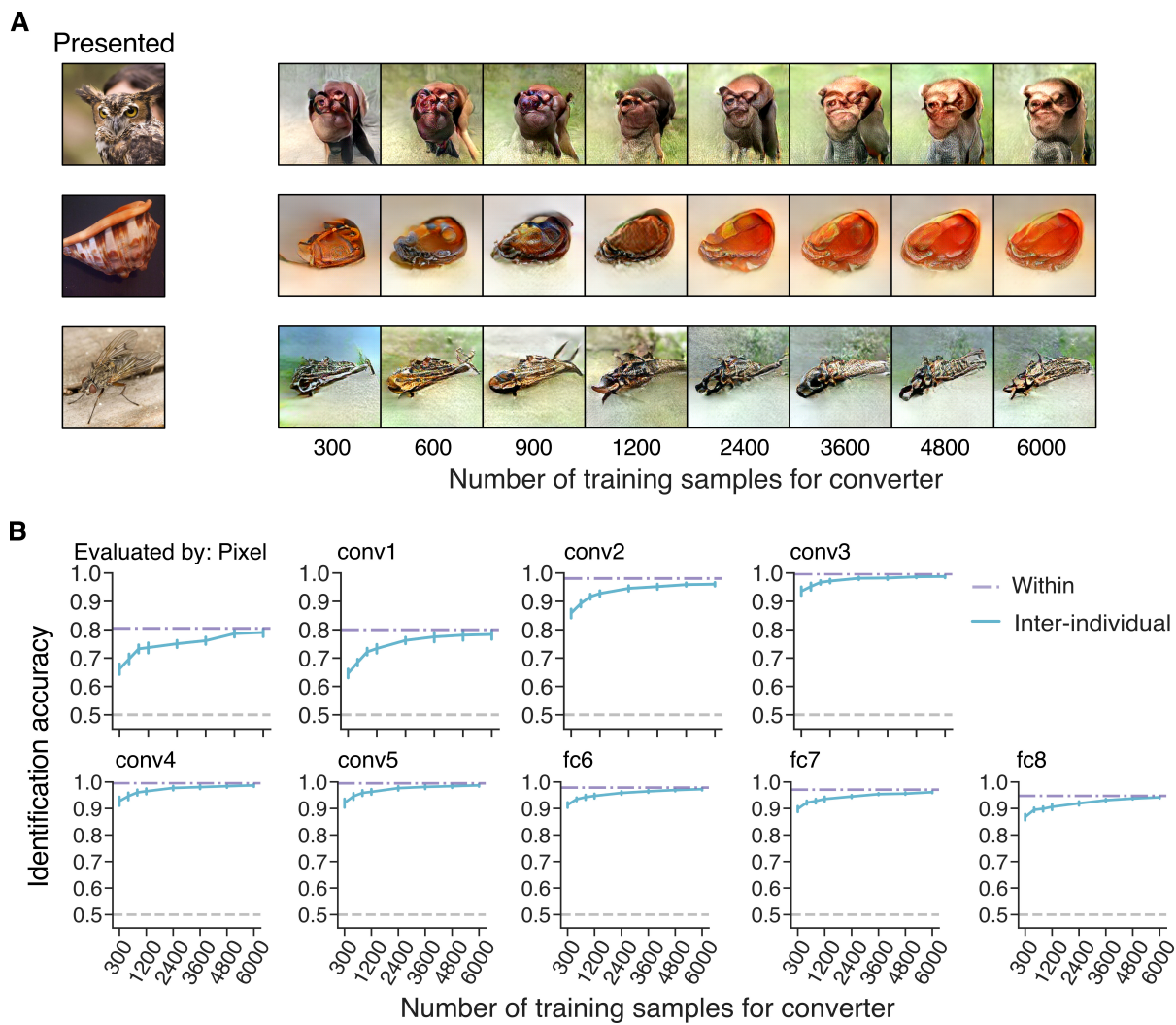
    {The effect of the number of training data for conversion.} 
    {\textbf{(A)} Reconstructed images using the VGG19 decoder. Reconstructed images were generated with a varying number of training data from the same pair (VC; source: Subject 1, target: Subject 2: See Supplementary Fig.14A for reconstructions using the CLIP\_ViT decoder).
    \textbf{(B)} Identification accuracy. Pairwise identification was performed using the pixel values and the extracted DNN feature values (AlexNet) from the reconstructions with a varying number of training data. The mean accuracy was calculated in each subject and condition (error bars, 95\% C.I. from 20 pairs; dashed lines, 50\% chance level; See Supplementary Fig.14B for identification accuracy using the CLIP\_ViT decoder). The results are shown with those from the Within condition.}
\section*{Introduction}
Individual differences in brain organization are observed at different scales, from macroscopic anatomy to fine-grained functional topography \cite{fischl2008cortical,van2004surface,van2005population,cox2003functional,guntupalli2016model,haxby2011common}. These individual differences may complicate and confound group-level analyses, and make it challenging to generalize individual-specific models to other individuals. While anatomical alignment can help mitigate anatomical differences \cite{fischl2008cortical,van2004surface,van2005population}, it does not fully account for the variability in functional topography across individuals, likely due to idiosyncratic neural representations at finer-grained scales \cite{cox2003functional,guntupalli2016model,haxby2011common}.

Functional alignment has been pivotal in functional magnetic resonance imaging (fMRI) research, addressing individual differences in functional topography. This approach involves presenting the same stimuli to different subjects and aligning brain activity patterns to make them similar or shared across individuals \cite{haxby2011common,haxby2020hyperalignment,yamada2015inter,chen2015reduced,bilenko2016pyrcca,guntupalli2016model,bazeille2021empirical,ho2023inter,thual2023aligning}. Functional alignment encompasses two primary strategies: pairwise alignments and template-based alignments. Pairwise alignments, such as neural code converters\cite{yamada2015inter, ho2023inter}, convert the brain activity pattern of one subject to another, representing the same content. Template-based alignments, like hyperalignment \cite{haxby2011common,chen2015reduced,guntupalli2016model,haxby2020hyperalignment}, create a common brain activity space for all subjects. However, functional alignment requires presenting identical stimuli (shared stimuli) to different individuals, resulting in paired brain activity data for model training. This requirement is often difficult to meet, hindering the widespread application of functional alignment techniques.

While shared stimuli in functional alignment ensure that brain activity patterns across subjects reflect the same stimulus content, the content can be more flexibly represented by a combination of elemental, latent features, such as image bases or deep neural network (DNN) features \cite{riesenhuber1999hierarchical,serre2007robust,mutch2008object,oliva2001modeling,lowe1999object,csurka2004visual,krizhevsky2012imagenet,simonyan2014very}. By extending the representation of content from entire images to these elemental features, functional alignment may encompass a broader range of stimuli beyond those that are explicitly shared. Recent advancements in brain decoding research have revealed that the contrasts of local image bases and the latent features from DNNs can be decoded from the brain activity patterns measured by fMRI \cite{miyawaki2008visual,horikawa2017generic}. Furthermore, these decoded features can be transformed into images by linear or deep generative networks \cite{miyawaki2008visual,shen2019end,shen2019deep,cheng2023reconstructing}, reconstructing perceptual experiences. These results pave the way for using latent features instead of shared stimuli to ensure content similarity in functional alignment. 

 We introduce a flexible pairwise functional alignment model using content loss-based optimization, which relies on the latent feature representation of stimulus content and does not require subjects to be exposed to shared stimuli. For each pair of subjects, we designate one as the target subject and the other as the source subject. First, we pre-train decoders to predict the latent features of images the target subject views from their measured brain activity \cite{horikawa2017generic}. Next, a neural code converter model is optimized to minimize the content loss between the latent features of the stimulus and those decoded from the converted brain activity patterns of the target subject (\figcite{fig:figureSuggestions}{A}). Importantly, this training approach for the converter does not rely on paired brain activity (shared stimuli), allowing the source and target subjects to originate from different datasets or without any shared stimuli. In the test stage, the trained converter is used to convert the source subject's brain activity in response to novel stimuli into the target brain space. The converted brain activity can be decoded into latent features, which are then used to generate images using a generator, reflecting the extent to which fine-grained visual information is preserved across the inter-individual conversion. This procedure can be performed across individuals (inter-individual image reconstruction) and even across measurement sites (inter-site image reconstruction), as illustrated in \figcite{fig:figureSuggestions}{B}. In this case, brain data measured at one site are converted to the brain space of a subject in another site, and then using the target site's decoder, feature decoding and image reconstruction can be performed.

\figureSuggestions
In this study, we first demonstrate that converters optimized using visual content can accurately convert brain activity patterns across subjects and capture fine-grained visual features for inter-individual image reconstruction, with performance comparable to converters optimized using paired brain activity. We then show that similar conversion accuracy and inter-individual image reconstruction quality can be obtained using either overlapping or non-overlapping stimuli between converter and decoder training, indicating that shared stimuli are unnecessary for our method. To further validate our approach, we construct inter-site neural converters across various datasets and demonstrate that inter-site image reconstructions faithfully reflect the viewed images, with quality approaching that of within-site reconstructions. An analysis of converters trained using different levels of image representations as visual content supports the notion that the use of multiple hierarchical visual features contributes to accurate neural code conversion. Moreover, we demonstrate that the converted brain activity is not specifically tailored to the specific readout and can be effectively read out by alternative decoding schemes. Finally, we show that images with recognizable object silhouettes can be reconstructed even when the converter is trained with limited amounts of data. These results demonstrate that our proposed model can perform functional alignment without the need for shared stimuli and can capture fine-grained features for inter-individual and inter-site image reconstruction.

\section{Results}
The inter-individual and inter-site neural code conversion analyses used three fMRI datasets focused on natural images and involved 68 subject pairs. The primary analysis was performed within the dataset introduced in our earlier studies \cite{shen2019deep,horikawa2022attention,ho2023inter}, referred to here as the “Deeprecon” dataset (see Methods: “Datasets”). This dataset comprises five subjects, each with 6,000 training samples (1,200 natural images with five repetitions), 1,200 test samples (50 natural images with 24 repetitions), and an additional 800 test samples for generalization evaluation (40 artificial images with 20 repetitions), totaling 20 subject pairs for analysis. To further evaluate the efficacy of our method, we extended our analysis to inter-site analysis, using the Deeprecon dataset along with two additional datasets: the Natural Scene Dataset (NSD) \cite{allen2022massive} and the THINGS dataset \cite{hebart2023things} (see Methods: “Datasets”). Within the NSD, we used two subjects, each provided with available 24,980 training samples (9,000 natural images with three repetitions) and 300 test samples (100 natural images with three repetitions). For the THINGS dataset, our analysis used two subjects, each having 8,640 training samples (8,640 natural images with one repetition) and 1,200 test samples (100 natural images with 12 repetitions). Consequently, 48 subject pairs were included in the inter-site analyses. Unless otherwise noted, all training samples in each subject were used in inter-individual or inter-site analyses as a representative case. For training neural code converters, we used the hierarchical DNN features from the VGG19 model \cite{simonyan2014very} (see Methods: “DNN models”) as the latent representation of stimulus content (\figcite{fig:figuretwo}{A, B}). In our decoding and reconstruction analyses, we used fMRI activity from the whole visual cortex (VC), and the test samples were averaged across repetitions for each image.

\figuretwo

\clearpage
\subsection{Neural code conversion} 
We first investigated whether brain activity patterns could be reliably aligned across subjects using a content loss-based converter. Conversion accuracy was evaluated by two metrics: (a) profile correlation, which is the Pearson correlation coefficient between the sequences of converted and measured responses of a single voxel to the 50 natural test images (\figcite{fig:figuretwo}{C}), and (b) pattern correlation, which calculates the Pearson correlation coefficient between the converted and measured voxel patterns for a test image (\figcite{fig:figuretwo}{D}). The correlation coefficients obtained were normalized against their noise ceilings to account for the noise present in fMRI brain responses across repeated measurements using the same stimulus \cite{hsu2004quantifying,lescroart2019human} (see Methods: “Evaluation of conversion accuracy”). To summarize the results, we averaged the profile correlation coefficients across all voxels and the pattern correlation coefficients across all images within each individual pair and each region of interest (ROI; see Methods: “Datasets”). For comparison, we also evaluated the brain loss-based neural code converter \cite{yamada2015inter,ho2023inter}(\figcite{fig:figuretwo}{B}; see Methods: “Methods for functional alignment”). The conversion analyses were performed using Deeprecon samples for each individual pair. 

\figcite{fig:figuretwo}{E, F} present the distributions of profile correlation coefficients and pattern correlation coefficients across all conversion pairs for different ROIs in the target brain space. Each data point in the figures corresponds to an individual pair. The mean profile correlation for the whole VC using content loss-based neural code converter was 0.51$\pm$0.05 (mean with 95\% confidence interval) for 20 individual pairs (\figcite{fig:figuretwo}{E} for group results; Supplementary Fig.1 left for individual pairs), whereas the mean pattern correlation for the VC was 0.62$\pm$0.08 for 20 individual pairs (\figcite{fig:figuretwo}{F} for group results; Supplementary Fig.1 right for individual pairs). The subareas exhibited distributions similar to those of the VC. The accuracy of the brain loss-based converter aligned with the results reported by Yamada et al.\cite{yamada2015inter} and Ho et al.\cite{ho2023inter}, while the content loss-based converter exhibited comparable performance across all examined visual subareas. These results indicate that converters optimized by visual content can also accurately convert fMRI activity patterns across subjects.

\subsection{Inter-individual DNN feature decoding}
We next investigated whether fine-grained feature representations of seen images were preserved in the converted fMRI activity patterns by a DNN feature decoding analysis \cite{horikawa2017generic}. Feature decoders were trained to predict DNN feature values of the stimuli from the target subject's fMRI activity patterns across the whole VC. These decoders were then tested on the converted brain activity to predict the DNN features of the test images (\figcite{fig:figurethree}{A}; see Methods: “DNN feature decoding analysis”). We evaluated the decoding accuracy using two metrics: (a) profile correlation, which is the Pearson correlation coefficient between the sequences of the decoded and true feature values across all test images for a DNN unit, and (b) pattern correlation, which is the Pearson correlation coefficient between the pattern of decoded features and the true feature patterns for a test image.  We further averaged the profile correlations across all DNN units and the pattern correlations across all test images within each layer. To provide a baseline for comparison, we calculated the decoding accuracy for the standard within-individual decoding, which predicts DNN features using decoders trained exclusively on data from the same subject, referred to as the “Within” condition. 

\figurethree
\figcite{fig:figurethree}{B, C} show the DNN feature decoding accuracy for the whole VC across three analytical conditions: Within, Brain loss, and Content loss. The content loss-based converters exhibit comparable decoding accuracies with those obtained through the Within decoding, with consistently similar trends across various DNN layers. The content loss-based converters show higher accuracy than those employing brain loss regarding both profile correlation and pattern correlation. The results suggest that content loss-based converters effectively preserve the fine-grained representation of visual features and demonstrate an advantage in decoding DNN features over brain loss-based converters.

\subsection{Inter-individual visual image reconstruction}
We further reconstructed images using fine-grained DNN feature representations decoded from converted brain activity (\figcite{fig:figurethree}{A}; see Methods: “Visual image reconstruction”) and showed examples of the reconstructions from the whole VC (\figcite{fig:figurethree}{D, E}; see Supplementary Figs.2, 3 for the reconstructions of all pairs). We compared the reconstructions from three analytical conditions: Within, Brain loss,  and Content loss. The reconstructions obtained from Brain loss and Content loss conditions captured the essential characteristics and details of the presented images, with the visual objects being similarly recognizable to those in the Within reconstructions.

For a quantitative evaluation of reconstruction performance, we performed a pairwise identification analysis. This analysis used the pixel and DNN feature patterns of the reconstructed image to identify the true stimulus between two alternatives (see Methods: “Identification analysis”). We used the AlexNet model\cite{krizhevsky2012imagenet} to extract DNN feature patterns. The identification process was repeated with multiple false alternatives to obtain the accuracy for each reconstructed image. We then calculated the mean identification accuracy across all reconstructions for each individual pair, presenting these accuracies at the group level in \figcite{fig:figurethree}{F, G}. It was observed that both types of converters yielded lower identification accuracies compared to the Within reconstructions for both natural and artificial images. The content loss-based converter slightly outperformed the brain loss-based converter across all DNN layers for natural images and most layers for artificial images. Content loss-based converters consisting of different architectures showed similar performance (Supplementary Fig.4).


\subsection{The effect of stimulus overlap between converter and decoder trainings}
We have shown the results of inter-individual neural code conversion and image reconstruction when the training of the converter and decoder involves overlapping (shared) stimuli. However, it remains unclear whether we can perform neural code conversion without shared stimuli.To examine this, we randomly divided the training samples from the Deeprecon dataset into two distinct halves based on the categories of stimuli. The source subject were provided with 3000 training samples (600 images from 75 randomly selected categories out of 150 categories, with five repetitions of each image), and the target subject were given a different set of 3000 training samples (the remaining 600 images with five repetitions each), a condition we refer to as “Non-overlapping” (\figcite{fig:figurefour}{A}). This strategy was designed to prevent any pairing in their brain activity patterns. We then trained neural code converters for each individual pair under this condition and evaluated their conversion accuracy, inter-individual decoding accuracy, and image reconstruction. For comparison, we also performed the same analysis under an “Overlapping” condition, where the target subject used training samples the same as the source subject, resulting in overlapping stimuli between the converter and decoder trainings.

\figcite{fig:figurefour}{B} presents the conversion accuracy across the whole VC under two analytical conditions for all individual pairs. The mean profile correlation under the Non-overlapping condition was 0.41$\pm$0.05 (mean with 95\% confidence interval) for 20 individual pairs, whereas the mean pattern correlation was 0.56$\pm$0.06 for 20 individual pairs (see Supplementary Fig.5A, B for conversion accuracies across different ROIs). The converter and decoder trainings using non-overlapping stimuli show comparable conversion accuracy to those using overlapping stimuli. As shown in Supplementary Fig.5C, D, converters using non-overlapping stimuli show a similar decoding tendency across all DNN layers to those using overlapping stimuli, with accuracy levels approaching those observed in the Within decoding. The images reconstructed under two conversion conditions were recognizable and closely matched the presented stimuli as well as the within reconstructions (Supplementary Fig.5E). Quantitative evaluations of image reconstructions indicate that using either overlapping or non-overlapping stimuli between converter and decoder trainings results in similar identification accuracy (\figcite{fig:figurefour}{C}). These results demonstrated that the content loss-based converters can convert brain activity patterns across subjects without the need for overlapping (shared) stimuli, even between decoder and converter trainings.

\figurefour

\subsection{Inter-site neural code conversion and image reconstruction}
We extended our analysis to investigate the feasibility of inter-site neural code conversion, where converters were trained between source and target subjects from distinct sites. For this analysis, we used the Deeprecon dataset and two additional datasets: the NSD and the THINGS dataset. The source subject was selected from one of these three datasets, while the target subject was selected from a different dataset, resulting in 48 individual pairs with no stimuli shared between any pairs. Due to the absence of ground truth (measured fMRI activity) for evaluating conversion accuracy in the brain space, we opted to evaluate the inter-site neural code conversion through the decoding accuracy and image reconstruction from the converted brain activity.

The inter-site decoding shows a similar decoding tendency and comparable accuracy across all DNN layers compared to the Within decoding for test images from each dataset (Supplementary Fig.6). \figcite{fig:figurefive}{A-C} present examples of the inter-site image reconstruction (see Supplementary Figs.7–9 for the reconstructions of 48 individual pairs). The inter-site reconstructions captured the core characteristics of the presented images, including their shape, color, and textures, showcasing visual content similar to that of the Within reconstructions. A quantitative evaluation revealed that the identification accuracy of inter-site reconstructions was slightly lower but still comparable to Within reconstructions for test images from each dataset (\figcite{fig:figurefive}{D-F}). These results demonstrated that brain activity patterns can be converted across subjects from different datasets, preserving the fine-grained feature representations that enable inter-site image reconstruction. 

\figurefive

\subsection{Converters trained using the loss from different DNN layers}
To examine the significance of the use of multiple hierarchical DNN layers in converter training, we compared the conversion performance using the loss from different sets of DNN layers. \figcite{fig:figuresix}{A, B} compared the conversion accuracy between these converters, revealing that the loss from all layers results in the highest accuracy. However, it was observed that the conversion accuracies were lower when using loss from the lower and middle layers (conv1\_1 to conv5\_4). The higher layers, specifically fc6, fc7, and fc8, showed the worst conversion accuracy.

The decoding accuracy of the converters trained with the loss from all layers shows an advantage across all evaluated DNN layers (Supplementary Figs.10, 11). \figcite{fig:figuresix}{C} shows the images reconstructed by converters trained using the loss from different DNN layers. Images reconstructed from the “All layers” condition produced visual contents with high perceptual quality. In the case of converters trained using the loss from single DNN layers, specifically the lower and middle layers, the reconstructions also yielded recognizable visual contents. However, those trained with the higher layers were found to be less effective in capturing the details of the objects, leading to reconstructed images with lower perceptual quality. Quantitative evaluations presented in \figcite{fig:figuresix}{D} reflected this qualitative assessment. The reconstructions from the “All layers” condition achieved higher identification accuracy compared to those from each single-layer condition, ranging from conv4\_1 to fc8 layers. These results indicate the use of the loss from all DNN layers is important for accurate neural code conversion. 

\figuresix

\subsection{Reconstruction by converter and decoder trained with different DNNs}
To confirm whether the converted brain activity contains generalizable representations rather than those tailored for a specific readout (the VGG19 decoder), we examined the feasibility of decoding the converted brain activity using a different scheme. A different feature decoder was trained to predict CLIP\_ViT features \cite{radford2021learning}(see Methods: “DNN models”) of the stimuli from the target subject's fMRI activity patterns. This decoder was then tested on the converted brain activity obtained from VGG19-based converters. We further reconstructed images from the decoded CLIP\_ViT features using the same reconstruction method. 

Decoding CLIP\_ViT features from converted brain activity demonstrates accuracies similar to those of the Within decoding across all evaluated DNN layers (Supplementary Fig.12). \figcite{fig:figureseven}{A} shows examples of reconstructed images from these decoded CLIP\_ViT features (see Supplementary Fig.13 for the reconstructions of all pairs). While CLIP\_ViT features were not used during the converter's training, the reconstructions with the CLIP\_ViT decoder accurately reflected the visual object in the presented images. The reconstructed contents resemble those obtained with the VGG19 decoder, although the details differ slightly. The quantitative evaluation showed that reconstructions with the CLIP\_ViT decoder yielded identification accuracy approaching that of reconstructions with the VGG19 decoder across all evaluations (\figcite{fig:figureseven}{B}). These results suggest that the brain activity converted using our method is not tailored for the specific decoder and can be read out by different decoding schemes.

\figureseven

\subsection{Varying the number of training data for conversion}
We investigated the effect of training sample size for converters on image reconstruction quality. We varied the number of training samples for converters (trained with VGG19 features) at various levels: 300, 600, 900, 1,200, 2,400, 3,600, 4,800, and 6,000 training samples, with data collection time ranging from 40 minutes to approximately 13 hours. To evaluate the decoding from converted brain activity, we consistently used 6,000 samples from the target subject to train the VGG19 and CLIP\_ViT feature decoders.

\figcite{fig:figureeight}{A} shows the reconstructed images with the VGG19 decoder, using converters trained with varying sample sizes. Though the visual quality of the reconstructions diminished with smaller training sample sizes, converters trained with as few as 300 or 600 samples (around one hour) still yielded perceptible images. The identification accuracy increased with the number of training samples, gradually approaching the accuracy observed in the Within condition (\figcite{fig:figureeight}{B}). Similar results were obtained for the reconstruction using the CLIP\_ViT decoder (Supplementary Fig.14). These results indicate the feasibility of inter-individual image reconstruction with converters trained on limited data, achieving modest performance while decreasing the dependence on extensive fMRI data collection.

\figureeight
\clearpage
\section{Discussion}
Our study aimed to develop a functional alignment method that (1) eliminates the need for identical sets of stimuli (shared stimuli) to be presented to individuals during training and (2) captures the fine-grained visual representations of stimuli, enabling image reconstruction comparable to within-individual analyses. To achieve this, we proposed a flexible approach for training neural code converters using content loss-based optimization, which optimizes the converter parameters based on the visual content of stimuli as represented by latent features of a deep neural network (DNN) model, rather than relying on brain activity patterns paired by the same set of stimuli. 

Our proposed converter accurately converted a source subject's brain activity into a target subject's brain space, as evaluated by profile correlations and pattern correlations. The high perceptual quality of visual images reconstructed from the converted brain activity demonstrated that the converter captures fine-grained visual feature representations. Notably, training the converter with non-shared stimuli yielded comparable conversion accuracy and inter-individual image reconstruction performance, and successful reconstruction using a decoder from different datasets further confirmed the effectiveness of our method. The results of converters trained using the content loss from different DNN layers revealed that the use of multiple DNN layers was critical for achieving high conversion accuracy. The converted brain activity could be decoded into the features of another DNN model not used for converter training, to enable successful image reconstruction, indicating that the converter captures generalizable representations beyond the specific readout. Moreover, even with a limited amount of training data, images with recognizable object silhouettes could be reconstructed from the converted brain activity. 

We have demonstrated inter-site image reconstruction using the Deeprecon, THINGS, and NSD datasets. While all combinations of these datasets for conversion yielded reliable reconstructions, it should be noted that there are different characteristics among these datasets, particularly in the diversity of the stimuli. The Deeprecon dataset includes fMRI brain responses to both natural and artificial images and is carefully designed to prevent overlap in object categories between training and testing datasets. The NSD and THINGS datasets provide more extensive stimuli of natural scenes and object images. However, the diversity of NSD stimuli is limited in semantic and visual contents, with considerable overlap of these contents between training and test sets \cite{shirakawa2024spurious}. This overlap might introduce bias in inter-site reconstruction results, particularly with the NSD test set, because semantic and visual contents from the training data that overlap with the test data could be incorporated during the training of the converter. The use of multi-subject data in decoding may increase the risk of spurious prediction of “novel” content due to the growing overlap of semantic and visual contents between the training and test sets\cite{scotti2024mindeye2,quan2024psychometry,bao2024wills,gong2024mindtuner}.

\subsection{Content loss versus brain loss in functional alignment}
A major difference between our functional alignment method and the previous ones is that we optimized the converter based on the visual content of stimuli rather than on paired brain activity patterns. The optimization process is designed to ensure that the converted brain activity patterns can be decoded into latent feature representations that closely resemble the feature representations extracted from the stimulus. By performing the optimization within the visual content space, the content loss-based converter eliminated the need for paired brain activity patterns during the training stage. In contrast, conventional brain loss-based functional alignment methods \cite{yamada2015inter,bazeille2021empirical,ho2023inter} optimize within a brain space, aiming to minimize the discrepancy between measured brain activity patterns of different individuals when presented with the same stimulus. 

We used a brain loss-based converter trained with paired measured brain activity for the same stimulus for comparison. It is important to note that “brain loss” can be defined in various ways. For instance, brain loss can be measured using latent representations of brain data, and the alignment can be performed within the latent brain space rather than the innate brain space \cite{haxby2011common,chen2015reduced,guntupalli2016model,haxby2020hyperalignment,chen2023seeing,qian2023fmri,huo2024neuropictor}. Furthermore, brain loss is not limited to measured brain activity patterns. By employing subject-specific pre-trained encoders to predict brain activity patterns for non-shared stimuli, brain loss-based optimization can be applied to these predicted brain data in the same way as measured brain data \cite{wasserman2024functional}. 

These brain loss-based functional alignment methods may have common constraints. A brain's response to a stimulus comprises three components: a consistent stimulus-evoked response across individuals, an idiosyncratic stimulus-evoked response, and a noise component\cite{nastase2019measuring}. As a result, brain loss-based alignment methods might inadvertently incorporate idiosyncratic responses and noise, potentially limiting their effectiveness in capturing the consistent stimulus-evoked response. This could explain the inferior performance observed in \figcite{fig:figurethree}. In contrast, our proposed method may help mitigate the influence of idiosyncratic responses and noise components, leading to a more accurate representation of the consistent stimulus-evoked response across individuals.


Our content loss-based converter used hierarchical DNN features as latent representations for visual content. This method builds upon recent studies in brain decoding, which have identified parallels between the hierarchical representations in the human brain and the fine-grained features of DNNs \cite{horikawa2017generic}. By reconstructing visual images from converted brain activity, we demonstrated that the content loss-based converter indeed preserved fine-grained visual features (\figcite{fig:figurethree}{D-G}). Furthermore, we found that the results were robust across various model architectures for the content loss-based converter, including Nonlinear Multi-Layer Perceptron (MLP), Linear MLP, Residual MLP, and Vision Transformer (ViT) (Supplementary Fig.4). These findings support the effectiveness and robustness of using hierarchical DNN features as latent representations for visual content in neural code conversion.

Optimizing converters using content loss from multiple hierarchical DNN layers contributes to the improvement of neural conversion accuracy. However, when focusing on the quality of inter-individual image reconstructions, we found that using loss from single layers at either the lower or middle levels (specifically from conv1\_1 to conv4\_4) can match the performance of the loss from all layers (\figcite{fig:figuresix}{C, D}). Relying on single-layer loss from the higher DNN layers (fc6, fc7, and fc8) tends to degrade performance. This observation may be due to the substantial capacity of the lower and middle layers of DNN features to the image-level visual information, which is pivotal for reconstructing perceptually coherent images. Converters trained with higher DNN layers may overlook some fine-grained visual details while preserving more abstract information. These results suggest that integrating visual features across multiple levels is important for accurately converting brain activity patterns, preserving not only image-level information but also capturing the different levels of abstraction in visual processing.

It should be noted that latent representations for visual content are not limited to DNN features, but also include other visual features such as HMAX \cite{riesenhuber1999hierarchical,serre2007robust,mutch2008object}, GIST \cite{oliva2001modeling}, and the integration of scale-invariant feature transform (SIFT)\cite{lowe1999object} with the “Bag of Features” (BoF) approach \cite{csurka2004visual}. Visual features like DNN and HMAX features mimic the hierarchical structure of the human visual system, while GIST and SIFT + BoF are specifically designed to capture global scene properties and local image features, respectively. These visual features analyze visual contents at multiple levels and scales, and their representations have been reported to be statistically similar to visual cortical activity \cite{riesenhuber1999hierarchical,serre2007robust,cadieu2014deep,yamins2014performance,khaligh2014deep,gucclu2015deep,rice2014low,leeds2013comparing}. It remains to be seen which visual features are most appropriate as latent representations for visual content in the context of neural code conversion.

\subsection{Aligning brain space via content loss }
In our approach, brain activity patterns were indirectly aligned via the loss of latent representations of the visual content. This raises a concern that the converted brain activity could be specifically tailored to the particular form of content representation. However, we have found that although the converter was trained with VGG19 features \cite{simonyan2014very}, the converted brain activity could be decoded into CLIP\_ViT features \cite{radford2021learning}, which differ significantly from VGG19 features, resulting in faithfully reconstructed images (\sinfigcite{fig:figureseven}). This indicates that our converter does not simply convert brain activity to a format specially tailored for a specific DNN representation. Rather, it learns a generalizable brain activity representation interpretable by other decoding schemes. Although our current study focuses on image reconstruction, the converted brain activity could be used with other decoding schemes and other types of information and tasks, to examine the versatility of the conversion based on DNN representations.

The content loss-based converter employed a training approach similar to feature decoders, where it constituted additional layers to the target's decoder. However, our results suggest that it uniquely specialized in learning the statistical relationship between the source and target brain spaces in several key aspects. First, the converter effectively converted brain activity patterns across subjects, achieving conversion accuracy comparable to converters optimized by brain loss (\sinfigcite{fig:figuretwo}), highlighting its capability to align brain activity patterns. Second, as discussed above, the converted brain activity could be interpreted by another decoding scheme (\sinfigcite{fig:figureseven}), indicating that the converter learns a more generalizable mapping between brain spaces. Further, converter training required a substantially smaller amount of data for successful image reconstruction than feature decoder training (\sinfigcite{fig:figureeight}), suggesting a separate underlying mapping structure distinct from feature decoding.

\subsection{Inter-individual decoding without shared stimuli}

Our primary goal was to convert one individual's brain patterns to another's innate brain space without shared stimuli, allowing for use in downstream tasks including decoding and visual image reconstruction as a secondary goal. However, one can achieve inter-individual decoding without aligning brain activity patterns in the innate space. A multi-subject shared decoder can be trained with an alignment layer by optimizing with specific tasks \cite{zhou2024clip,liu2024see,scotti2024mindeye2,quan2024psychometry,bao2024wills,gong2024mindtuner}. This method can implicitly create a shared latent space for brain data. Similar approaches to handling non-shared stimuli have been applied to decoding in other neuroimaging modalities, such as magnetoencephalography (MEG) \cite{defossez2023decoding,benchetrit2023brain}, electroencephalography (EEG) \cite{defossez2023decoding}, and electrophysiological neural recordings \cite{azabou2024unified}.

While these inter-individual decoding studies have shown promising results for particular decoding tasks, the alignment of brain activity patterns is performed implicitly with decoder training, and alignment in the innate space is not their goal. Thus, the potential for converting brain activity for general purposes is not carefully examined. In contrast, our neural code conversion method explicitly generated the brain data in the subjects' innate brain space, allowing for explicit evaluation of the conversion across brain spaces. The successful decoding and image reconstruction (\figscite{fig:figurethree}{fig:figurefour}) obtained by applying the target’s original model to the converted brain activity further underscore the effectiveness of our method. The flexibility in converting brain data and seamless compatibility with existing data analysis pipelines potentially pave the way for scalable data analyses.

\subsection{Scalable data analysis and applications}
Our method provides a promising tool for scalable neural data analyses. Conventional within-site analyses\cite{haxby2011common,chen2015reduced,haxby2020hyperalignment,ho2023inter,yamada2015inter,bilenko2016pyrcca,van2018modeling,scotti2024mindeye2,quan2024psychometry,bao2024wills,gong2024mindtuner}, are often conducted under homogeneous conditions, such as the same scanner, identical experimental design, and subjects from a single site, which may restrict the generalizability of the results. In contrast, our method enabled the conversion of brain data across different datasets and demonstrated that inter-site converted brain data can achieve image reconstructions with performance comparable to within-site reconstructions (\sinfigcite{fig:figurefive}). This approach potentially expands the reuse of brain data from various publicly available datasets, and enables scalable data analysis and modeling that transcend institutional and geographical barriers. 

The neural code converter has the potential to reduce both the economic costs and time investments required for data collection. It can be applied to decrease the amount of data required for model training on novel subjects by using training data from other individuals. Since fMRI data collection is time-consuming and expensive, the efficient use of data is crucial in neuroimaging research. For example, in a block design experiment where each stimulus image lasts for 8 s, collecting 6000 fMRI training samples typically requires around 13 hours. With our neural code converter, fewer training data samples are needed; for instance, only 600 samples (around one hour) might be sufficient for inter-individual decoding analyses, such as inter-individual image reconstruction (\sinfigcite{fig:figureeight}). This approach enables more inclusive studies that can accommodate a wider range of subjects, including those who may have difficulties participating in long scanning sessions, such as children, elderly populations, or individuals with certain disabilities. 

The content loss-based neural code converter may flexibly deal with brain data from more complex cognitive tasks and different modalities beyond visual perception. If the features or parameters of such tasks can be reliably decoded, the content loss could be extended beyond visual features, allowing for the alignment of brain data with viable time courses of behavior and cognitive states. Additionally, brain data from different modalities can be flexibly aligned using our method. For instance, high spatial resolution fMRI data can be converted into representations in other modalities, such as high temporal resolution EEG or MEG, enabling inter-modality decoding.

The neural code converter may also help design high-resolution brain stimulation \cite{polania2018studying,lozano2019deep,reutsky2013holographic}. As brain activity patterns are converted into the subject’s innate brain space while preserving fine-grained information, they can be used to design high-resolution brain stimulation patterns and to induce mental content similar to that of the source subject. This enables the transmission of mental content between two individuals, providing a foundation for brain-to-brain communication.

\section{Methods}
\subsection{Datasets}
We reanalyzed three existing datasets published previously and publicly available \cite{shen2019deep,horikawa2022attention,ho2023inter,allen2022massive,hebart2023things}. In this study, we used nine subjects from these three datasets. Subjects 1–5 are the individuals from the Deeprecon dataset \cite{shen2019deep,horikawa2022attention,ho2023inter}. Subjects 6 and 7 correspond to subjects 1 and 2 in the THINGS dataset \cite{hebart2023things}, and Subjects 8 and 9 correspond to subjects 1 and 2 from the NSD\cite{allen2022massive}. We used the preprocessed fMRI data released by these datasets \cite{shen2019deep,horikawa2022attention,ho2023inter,hebart2023things,allen2022massive}, with the ROIs they defined. A brief overview of the data we used is provided below.

\subsection{\textit{Deeprecon dataset}}
\noindent\textbf{Subjects.} Five healthy subjects (four males and one female; age range, 25–36 years) with normal or corrected-to-normal vision participated in the experiment. 

\noindent\textbf{Stimuli.} The natural image stimuli were selected from 200 representative categories in the ImageNet dataset (2011, fall release)\cite{deng2009imagenet}. The natural training images were 1,200 images taken from 150 object categories, and the natural test images were 50 images taken from the remaining 50 object categories \cite{horikawa2017generic,horikawa2019characterization}. The artificial test image stimuli consisted of 40 combinations of five shapes (square, small frame, large frame, plus sign, and cross sign) and eight colors (red, green, blue, cyan, magenta, yellow, white, and black).  

\noindent\textbf{Experimental paradigm and data acquisition.} The subjects were asked to fixate on the central fixation spot of the screen and to click a button when two sequential blocks presented the same image. Each presentation of an image lasted for 8 s in a stimulus block. A 3T scanner was used to collect whole-brain functional MRI data with 2-mm isotropic resolution and 2-s repetition time. fMRI signals were measured while subjects each viewed 1,290 visual images (8,000 trials) over the course of 15–20 scan sessions.

\noindent\textbf{MRI data preprocessing.} The fMRI data was preprocessed with the FMRIPREP pipeline\cite{esteban2019fmriprep}. The BOLD time series were temporally shifted by 4 s to account for hemodynamic delays and then regressed for nuisance variables. The data samples were finally despiked to reduce extreme values (beyond $\pm$3 SD for each run) in the time series and averaged within each 8-s trial (four volumes). The brain regions V1, V2, V3, and V4 were demarcated using a standard retinotopy experiment \cite{engel1994fmri,sereno1995borders} in each subject's native brain space. The higher visual cortex (HVC) was defined by conventional functional localizers \cite{kourtzi2000cortical,kanwisher1997fusiform,epstein1998cortical}. The whole visual cortex (VC) was defined as the combination of the regions V1, V2, V3, V4, and HVC.

\subsection{\textit{THINGS dataset}}
\noindent\textbf{Subjects.} Three healthy subjects (one male and two females; mean age: 25.33 years) with normal or corrected-to-normal vision participated in the experiment. 

\noindent\textbf{Stimuli.} The image stimuli were taken from the THINGS object concept and image database \cite{hebart2019things}. The training images were 8,640 images taken from 720 representative object concepts, with the first 12 examples per concept, and the test images were 100 separate images taken from the remaining THINGS images. 

\noindent\textbf{Experimental paradigm and data acquisition.} The subjects were asked to keep their eyes on the fixation spot of the screen and report the presence of a catch image with a button press on a fiber-optic diamond-shaped button box. Each image was presented for 0.5 ms, followed by 4 s of eye fixation without image stimuli. A 3T scanner was used to collect whole-brain functional MRI data with 2-mm isotropic resolution and 1.5-s  repetition time. fMRI signals were measured while subjects each viewed 8,740 unique visual images (11,040 trials) over the course of 15–16 scan sessions. 

\noindent\textbf{MRI data preprocessing.} The fMRI data was preprocessed using FMRIPREP, followed by ICA denoising. Then, fMRI single-trial responses were estimated by fitting a general linear model (GLM) on the preprocessed fMRI time series. The ROIs were defined based on retinotopic mapping \cite{engel1994fmri,sereno1995borders} and functional localizer experiments \cite{kourtzi2000cortical,kanwisher1997fusiform,epstein1998cortical}, with VC as the combination of all visual areas.

\subsection{\textit{Natural Scene Dataset (NSD)}}
\noindent\textbf{Subjects.} Eight healthy subjects (two males and six females; age range, 19–32 years) with normal or corrected-to-normal vision participated in the experiment.

\noindent\textbf{Stimuli.} The image stimuli were sourced from the 80 COCO categories within Microsoft’s COCO image database\cite{lin2014microsoft}. The training images comprised 9,000 images that were mutually exclusive across subjects, and the test images consisted of 100 special images taken from 1,000 images that were shared across subjects. 

\noindent\textbf{Experimental paradigm and data acquisition.} Subjects fixated centrally and performed a long-term continuous recognition task on the images. Images were presented for 3 s with 1-s gaps in between images. Scanning was conducted at 7T using whole-brain gradient-echo EPI at 1.8-mm resolution and 1.6-s repetition time. fMRI signals were measured while subjects each viewed 9,000–10,000 distinct natural scenes (22,000–30,000 trials) over the course of 30–40 scan sessions. 

\noindent\textbf{MRI data preprocessing.} The fMRI data were preprocessed by performing interpolation to correct for slice time differences and head motion. The GLM analysis was then used to estimate fMRI single-trial responses. Both the ROIs derived from atlases and those manually defined based on data from each subject are provided. In our analysis, the VC was generated from surface-based representations of the data using the HCP\_MMP1 atlas \cite{glasser2016multi}.

\subsection{Methods for functional alignment}

\subsection{\textit{Content loss-based neural code converter}}
The content loss-based neural code converter for each pair of subjects uses a nonlinear Multi-Layer Perceptron (MLP) to predict the brain activity patterns of one subject (target) from the brain activity patterns of another subject (source). It has two hidden layers with instance normalization and ReLU functions, and the number of units in each hidden layer is half that of the input layer. The converter $\Phi$ takes a source subject's brain activity pattern $\mathbf{x}_i\in\mathbb{R}^m$ consisting of $m$ voxels' values, and predicts the target subject's brain activity pattern as $\Phi(\mathbf{x}_i)\in\mathbb{R}^n$, consisting of $n$ voxels' values. Then, a pre-trained target decoder takes the converted brain activity pattern $\Phi(\mathbf{x}_i)$ and predicts the decoded feature pattern of the image stimulus as $\mathbf{d}_{il}= \mathbf{W}_{l}\Phi(\mathbf{x}_i)+\mathbf{b}_{l}$, where $\mathbf{d}_{il}\in\mathbb{R}^{d_l}$ is the decoded feature pattern consisting of $d_l$ units' values in the $l$-th DNN layer for the $i$-th image stimulus. $\mathbf{W}_l\in\mathbb{R}^{d_l\times n}$ and $\mathbf{b}_l\in\mathbb{R}^{d_l}$ are the decoding matrix and bias vector, respectively, of the pre-trained decoder.
The converter is optimized to make decoded features from the converted brain activity similar to the true features of the source subject’s stimulus. It is trained to minimize the objective function
\begin{equation}
	\mathcal{L}(\Phi) = \sum_i^N{}\sum_l^{L}\eta_{l}||\mathbf{v}_{il}-(\mathbf{W}_l\Phi(\mathbf{x}_i)+\mathbf{b}_l)||^2,
\end{equation}
where $\mathbf{v}_{il}\in\mathbb{R}^{d_l}$ represents true units' values in the $l$-th DNN layer for the $i$-th image stimulus, $N$ is the number of training samples, and $L$ is the number of DNN layers. $\eta_{l}$ is the parameter that weighs the contribution of the $l$-th layer, which is set to be $1/||\mathbf{v}_{il}||^2$.

We use the nonlinear MLP as the default architecture for the content loss-based converter. To verify the consistency of the conversion results across different converter architectures, we also used several alternatives: a Linear MLP with two hidden layers, a Residual MLP that integrates a nonlinear MLP with three residual blocks\cite{he2016deep}, and a Vision Transformer (ViT) that modifies the input and output layer of the original architecture used by Beyer et al.\cite{beyer2022better}.

The converter training was resolved through an iterative process. Each iteration involves a stochastic decoding strategy applied to all VGG19 layers (see Methods: “DNN models”). Specifically, during each iteration, within each convolutional layer, a feature map was chosen at random, and all its units were subjected to the decoding process. Due to the much fewer units in the fully connected layers, all units from these layers were decoded in each iteration. 1,024 iterations were used to ensure the comprehensive involvement of all DNN units from every layer of the image stimulus during the training stage. 

\subsection{\textit{Brain loss-based neural code converter}}
The brain loss-based neural code converter consists of a set of regularized linear regression models \cite{ho2023inter}. It takes a source subject's brain activity pattern $\mathbf{x}_i\in\mathbb{R}^m$ consisting of $m$ voxels' values, and predicts the target brain activity pattern as $\hat{\mathbf{y}_i}= \mathbf{M}\mathbf{x}_i+\mathbf{c}$, consisting of $n$ voxels' values. $\mathbf{M}\in\mathbb{R}^{n\times m}$ is the conversion matrix, and $\mathbf{c}\in\mathbb{R}^n$ is the bias vector. The converter is trained to minimize the objective function
\begin{equation}
	\mathcal{L}(\mathbf{M},\mathbf{c}) =\sum_i^N{}||\mathbf{y}_{i}-(\mathbf{M}\mathbf{x}_i+\mathbf{c})||^2 + \lambda||\mathbf{M}||^2_{F},
\end{equation}
where $\mathbf{y}_{i}$ is the measured target subject's brain activity pattern for the $i$-th sample, $N$ is the number of training samples, $\lambda$ is the regularization parameter, and $||\cdot||_{F}$ represents the Frobenius norm.

\subsection{DNN models}
We used the VGG19 DNN model \cite{simonyan2014very} implemented using the Caffe library \cite{jia2014caffe} in the converter training and DNN feature decoding analysis. This model is pre-trained for the 1,000-class object recognition task using the images from ImageNet \cite{deng2009imagenet}(the pre-trained model is available from \url{https://github.com/BVLC/caffe/wiki/Model-Zoo}). The model consists of 16 convolutional layers and three fully connected layers. All the input images to the model were rescaled to 224$\times$224 pixels. Outputs from individual units before rectification were used. The number of units in each layer is as follows: conv1\_1 and conv1\_2, 3,211,264; conv2\_1 and conv2\_2, 1,605,632; conv3\_1, conv3\_2, conv3\_3, and conv3\_4, 802,816; conv4\_1, conv4\_2, conv4\_3, and conv4\_4, 401,408; conv5\_1, conv5\_2, conv5\_3, and conv5\_4, 100,352; fc6 and fc7, 4,096; and fc8, 1,000.

For the evaluation of reconstructed images, we used another DNN model, the AlexNet \cite{krizhevsky2012imagenet} implemented using the Caffe library to extract DNN features from the reconstructed images and the presented image. This model is also pre-trained with images in ImageNet to classify 1,000 object categories (available from \url{https://github.com/BVLC/caffe/tree/master/models/bvlc\_alexnet}). The model consists of five convolutional layers and three fully connected layers. All the input images to the model were rescaled to 224$\times$224 pixels. The number of units in each layer is as follows: conv1, 290,400; conv2, 186,624; conv3 and conv4, 64,896; conv5, 43,264; fc6 and fc7, 4,096; and fc8, 1,000.

To examine different decoding schemes, we used the CLIP\_ViT model\cite{radford2021learning} implemented using the PyTorch library \cite{paszke2019pytorch}. This model is pre-trained on diverse image-text pairs to link images with text descriptions for a variety of visual tasks without task-specific training (available from \url{https://github.com/openai/CLIP}). All the input images to the model were rescaled to 224$\times$224 pixels. The number of units in each layer for the image encoder is as follows: conv1, 37,632; from transformer\_resblocks0 to transformer\_resblocks11, both the attn\_output and mlp layers in each block, 38,400; ln\_post, 768; and model\_output, 512.

\subsection{Evaluation of conversion accuracy}
We evaluated the conversion accuracy using two metrics: profile correlation and pattern correlation. Profile correlation is the Pearson correlation coefficient between the sequences of converted and true voxel responses to the 50 natural test images. Pattern correlation, which calculates the Pearson correlation coefficient between the converted and measured voxel patterns for a test image. Repeated measures of the brain responses to an identical stimulus in fMRI data are subject to measurement noise, which impacts the evaluation of conversion accuracy. To address this issue, we performed the noise ceiling estimation \cite{lescroart2019human,hsu2004quantifying}. The noise ceiling is calculated by averaging the correlation coefficients between repeated responses to identical stimuli, which reflects the maximum performance that the converter model can achieve given measurement noise in fMRI data. We excluded samples or voxels if their noise ceilings fell below a specified threshold (the 99th percentile of the distribution from random pairs). The obtained correlation coefficients were then normalized by dividing the raw values by their respective noise ceilings. For each individual pair and each ROI, the normalized profile correlation coefficients for all voxels were averaged to calculate the mean conversion accuracy (profile), while the normalized pattern correlation coefficients for all test stimuli were averaged to calculate the mean conversion accuracy (pattern). In the DNN feature decoding and image reconstruction analyses, we retained all voxels to avoid potential information leakage.

\subsection{DNN feature decoding analysis}
We used a ridge linear regression model as a DNN feature decoder. This model predicts the feature values of the stimulus, given an fMRI activity pattern evoked by the stimulus. We normalized both the feature values and voxel responses before model training and used a voxel selection procedure. This procedure involved calculating the Pearson correlation coefficients between sequences of voxel responses and feature values for all voxels. The 500 voxels exhibiting the highest correlations were selected for training. We set the ridge regularization parameter to 100 to enhance model robustness. The feature decoding analysis is detailed in the studies by Horikawa and Kamitani\cite{horikawa2017generic,horikawa2022attention}, and Shen et al.\cite{shen2019deep}. For testing the trained decoders, we used the average fMRI pattern over repetitions to improve the signal-to-noise ratio of the fMRI signal. We evaluated the decoding accuracy using profile correlation and pattern correlation. Profile correlation is the Pearson correlation coefficient between the sequences of the decoded and true feature values for the test images, while pattern correlation calculates the Pearson correlation coefficient between the pattern of decoded features and the true feature patterns for a test image. For each individual pair or each subject (Within), the profile correlation coefficients for all DNN units were averaged to calculate the mean decoding accuracy (profile), while the pattern correlation coefficients for all test stimuli were averaged to calculate the mean decoding accuracy (pattern). These mean accuracies were used as data points in group analysis.

\subsection{Visual image reconstruction}
The reconstruction method used in this study was extended from our original deep image reconstruction study \cite{shen2019deep}. The pixel values of an input image were optimized to make its image features match the decoded features from brain activity. Following Shen et al.\cite{shen2019deep}, we used the feature values before the rectification operation from eight layers (conv1-5 and all fully connected layers). We applied the same natural image prior and extended the loss function by adding a DISTS (Deep Image Structure and Texture Similarity) loss component \cite{Muraki2024}, which leverages spatial characteristics of feature maps to improve the detail of image reconstructions. Given the decoded features from multiple layers, an image was reconstructed by solving the following optimization problem:
\begin{equation}
	\mathbf{z}^* = \mathop{\arg\min}\limits_{\mathbf{z}} \left( \mathcal{L}_{\mathrm{mse}}(\mathbf{z}) + \lambda_{\mathrm{tex}} \mathcal{L}_{\mathrm{tex}}(\mathbf{z}) + \lambda_{\mathrm{str}} \mathcal{L}_{\mathrm{str}}(\mathbf{z})\right),
\end{equation}
where $\mathbf{z}$ is the latent vector, and $\mathcal{L}_{\text{mse}}(\mathbf{z})$ is the loss originally used in Shen et al.\cite{shen2019deep}:
\begin{equation}
	\mathcal{L}_\mathrm{mse}(\mathbf{z}) = \sum_{l}^{L} \gamma_l \left\| \Psi_l(G(\mathbf{z})) - \mathbf{u}_{l} \right\|^2.
\end{equation}
$G$ is the deep generator network (DGN) to enhance the naturalness of the image \cite{nguyen2016synthesizing}, and the reconstrued image is obtained as $G(\mathbf{z}^*)$. $\Psi_l$ is the function that maps the image to the DNN feature vector of the $l$-th layer. $\mathbf{u}_l\in\mathbb{R}^{P_l\times Q_l\times K_l}$ represents the decoded DNN feature vector of the image at the $l$-th layer, where $P_l$, $Q_l$, and $K_l$ denote the width, height, and number of channels of the feature maps in the $l$-th layer, respectively. $\gamma_{l}$ is the parameter that weights the contribution of the $l$-th layer and is set to be $1/||\mathbf{u}_{l}||^2$. $L$ is the number of DNN layers. $\mathcal{L}_{\mathrm{tex}}(\mathbf{z})$ is the texture similarity loss, and $\mathcal{L}_{\mathrm{str}}(\mathbf{z})$ is the structure loss; together, they constitute the DISTS loss.  $\lambda_{\text{tex}}$ and $\lambda_{\text{str}}$ serve as the coefficients of weights for the texture and structural similarity losses, respectively.

If we denote $\hat{\mathbf{u}}_{l}=\Psi_l(G(\mathbf{z}))\in\mathbb{R}^{P_l\times Q_l\times K_l}$, then the texture similarity $\mathcal{L}_{\mathrm{tex}}(\mathbf{z})$ and $\mathcal{L}_{\mathrm{str}}(\mathbf{z})$ are defined as follows:
\begin{equation}
	\mathcal{L}_{\mathrm{tex}}(\mathbf{z}) = -\sum^L_{l} \alpha_{l}\frac{1}{K_l} \sum_{k} \frac{
		\mu_k(\mathbf{u}_{l})  \mu_k(\hat{\mathbf{u}}_{l}) + \epsilon}   {\mu_k(\mathbf{u}_{l})^2 + \mu_k(\hat{\mathbf{u}}_{l})^2 + \epsilon},
\end{equation}

\begin{equation}
	\mathcal{L}_{\mathrm{str}}(\mathbf{z}) = -\sum^L_{l} \beta_{l}\frac{1}{K_l} \sum_{k} \frac{
		\delta_k(\mathbf{u}_{l},\hat{\mathbf{u}}_{l}) + \epsilon}   {\delta_k(\mathbf{u}_{l})^2 + \delta_k(\hat{\mathbf{u}}_{l})^2 + \epsilon},
\end{equation}
where
\begin{equation}
	\mu_k(\mathbf{u}_l) = \frac{1}{P_l Q_l}\sum_{p_l,q_l}\mathbf{u}_{p_l,q_l,k_l},
\end{equation}
\begin{equation}
	\delta_k({\mathbf{u}_l}) = \frac{1}{P_lQ_l}\sum_{p_l,q_l}({\mathbf{u}}_{p_l,q_l,k_l}-\mu_k(\mathbf{u}_l))^2,
\end{equation}
\begin{equation}
	\delta_k(\mathbf{u}_l, \hat{\mathbf{u}}_l) = \frac{1}{P_l Q_l}\sum_{p_l,q_l}{\mathbf{u}}_{p_l,q_l,k_l}{\hat{\mathbf{u}}}_{p_l,q_l,k_l}-\mu_k(\mathbf{u}_l)\mu_k(\hat{\mathbf{u}}_l).
\end{equation}
Here, $\mathbf{u}_{p_l,q_l,k_l}\in\mathbb{R}$ is a notation meaning the activity value of the $(p, q, k)$-th element for the $l$-layer. The $\alpha_l$ and $\beta_l$ are hyperparameters that signify the weight assigned to each layer, which were tuned using data from a subject excluded from the result analyses. A small positive constant $\epsilon$ is included to avoid numerical instability when the denominator is close to zero. We solved the optimization problem using stochastic gradient descent with momentum with 200 iterations.

\subsection{Identification analysis}
We employed identification analysis to quantify the accuracy of image reconstruction. This approach involved the identification of the presented image out of two alternatives based on the similarity of image features, including pixel values and DNN features. We reshaped the features of the reconstructed image into a one-dimensional feature vector and used this vector for comparison against the true feature vector of the presented image, as well as the false alternative of another image. An identification was considered correct when the correlation coefficient of the reconstructed image's feature vector was greater with the true feature vector than with the false alternative. This procedure was repeated for multiple false alternatives per reconstruction. We then defined the identification accuracy for a reconstructed image as the ratio of correct identifications. For each individual pair or each subject (Within), we calculated the mean identification accuracy by averaging the identification accuracies of all reconstructed images, which was used as a data point for group analysis.

\subsection{Group statistics}
We performed group-level statistical inference by showing the group means and confidence intervals in the figures. For DNN feature decoding and image reconstruction in within-individual analyses, each data point represents the mean accuracy for a subject. The mean accuracies from all subjects were used to calculate the group mean and its 95\% confidence interval. For conversion accuracy, DNN feature decoding, and image reconstruction in inter-individual analyses, each data point represents the mean accuracy corresponding to a unique pair of subjects.  Due to potential correlations between pairs involving the same subject (dyadic dependency), we applied bootstrapping to these dyadic data points to calculate the group mean and 95\% confidence interval. We performed bootstrap sampling separately on the source and target IDs, obtaining data points of the original sample size while ensuring that pairs where the source and target IDs were the same were excluded. The obtained data points were then used to calculate the mean. This process was repeated 1000 times to generate bootstrap replicates, from which we calculated the 95\% confidence interval. 

\section{Data Availability}
The experimental data that support the findings of this study are available publicly (Deeprecon dataset: \url{https://openneuro.org/datasets/ds001506/versions/1.3.1} for Subject 1-3, \url{https://openneuro.org/datasets/ds003430/versions/1.2.0} for the dataset of training natural-image session for Subject 4 and 5, and \url{https://openneuro.org/datasets/ds003993/versions/1.0.0} for the dataset of test natural-image and artificial-image sessions for Subject 4 and 5;
THINGS dataset: \url{https://openneuro.org/datasets/ds004192/versions/1.0.5}; 
Natural Scene Dataset: \url{https://naturalscenesdataset.org}). 

\section{Code Availability}
The code that supports the findings of this study can be found in our depository.

\bibliography{sample}

\section{Acknowledgements}
We thank our laboratory team, especially Yoshihiro Nagano, Ken Shirakawa, Eizaburo Doi, and Hideki Izumi, for their invaluable comments and suggestions on the manuscript. This study was conducted using the MRI scanner and related facilities of Institute for the Future of Human Society, Kyoto University. Funding was supported by Japan Society for the Promotion of Science (KAKENHI grant JP20H05705 and JP20H05954 to Y.K.), the New Energy and Industrial Technology Development Organization (Grant Number JPNP20006 to Y.K.), Guardian Robot Project, RIKEN, and Japan Science and Technology Agency (CREST grant JPMJCR18A5 and JPMJCR22P3 to Y.K.).

\section{Author contributions}
Conceptualization, H.W. and Y.K.; Methodology, H.W., J.K.H., Y.M., M.T., and Y.K.; Software: H.W. and S.C.A.; Formal Analysis, H.W.; Investigation, H.W., J.K.H., F.L.C; Writing–original draft: H.W. and Y.K.; Writing–review and editing: H.W., J.K.H., F.L.C, S.C.A., M.T., and Y.K.; Visualization, H.W. and Y.K.; Supervision: Y.K.; Funding Acquisition, Y.K.

\section{Competing interests}
The authors declare no competing interests.

\section{Supplementary information}
Supplementary Figures 1–14.

\noindent Supplementary Videos 1-3.

\end{document}